\documentclass{article}
\usepackage[margin=2cm]{geometry}  % Adjust the margin here
\usepackage{url}

\usepackage[
    backend=biber,
    style=numeric,
    citestyle=numeric,
    doi=false,
    isbn=false,
    url=false,
    maxnames=50,
    firstinits=true]{biblatex}
\usepackage{authblk}
\usepackage{physics}

\usepackage{graphicx} % Required for inserting images
\usepackage[hidelinks]{hyperref}
\usepackage{xcolor}
\usepackage{caption}
\usepackage{subcaption}
\usepackage{amsmath, amssymb, amsthm}
\usepackage{mathtools} % extension package to amsmath
\usepackage{soul}
\usepackage{siunitx} % for si units
\usepackage{enumerate}% http://ctan.org/pkg/enumerate
\usepackage{enumitem}% http://ctan.org/pkg/enumitem
\usepackage{float} % for placing figures
\usepackage{tikz} %for tikz images such as grids for flows
\usetikzlibrary{matrix}

\theoremstyle{definition}

\title{Influence of a generative parameter on the mechanical performance of
topological interlocking assemblies of a hexagonal block}
\date{}
\author[1,2]{Lukas~Schnelle}
\affil[1]{Chair of Algebra and Representation Theory, RWTH Aachen University, Pontdriesch 10-16, 52062 Aachen, Germany}
\affil[2]{Institute of Structural Mechanics and Lightweight Design, RWTH Aachen University, W\"ullnerstr. 7, 52062 Aachen, Germany}

\author[1]{Meike Weiß}

\author[1]{Reymond~Akpanya}
\author[2]{Kai-Uwe~Schr\"oder}
\author[1]{Alice~C.~Niemeyer}

\begin{document}

\maketitle

\begin{abstract}

A topological interlocking assembly is an arrangement of blocks, where all blocks are kinematically constrained by their neighboring blocks and a fixed frame. This concept has been known for a long time, attracting recent interest due to its advantageous mechanical properties, such as reusability, redundancy and limited crack propagation. New mathematical methods enable the generation of vast numbers of new topologically interlocking blocks. A natural next question is the quantification of the mechanical performance of these new blocks. We conduct a numerical study of topological interlocking assemblies whose blocks are constructed based on the hexagonal grid. By varying a design parameter used in the generation of these blocks, we study its influence on the structural performance of the entire assembly. The results improve our understanding of the link between the block parameters and the mechanical performance. This enhances the ability to custom design blocks for certain mechanical requirements of the topological interlocking assemblies.

\end{abstract}

\section{Introduction}

Constructing large modular structures from identical blocks has been a longstanding practice, particularly in civil engineering. Traditionally, such structures are composed of bricks which are cuboids, one of the simplest geometric forms. While straightforward, this approach often requires adhesive bonding between blocks, thereby introducing additional points of potential failure and complicating the processes of recycling and reuse. A promising alternative is to employ block geometries that admit the construction of so called \emph{topological interlocking assemblies} (TIAs). A TIA is an arrangement of blocks such that, once a subset of these blocks (called the \emph{frame}) is fixed, the remaining blocks are kinematically constrained solely through their contact with neighboring blocks and the frame.

In this paper, a new topological interlocking block, called the \emph{starfish block}, is introduced. By employing a design parameter, we are able to construct different versions of this block. This enables a comprehensive study of the mechanical performance of the resulting assemblies, derived from variations in the described design parameter. Here, we examine FEM simulations to study the interaction between the
blocks in the planar assemblies under transverse loading.

\subsection{Advantages and applications}

One of the main advantages of using a block admitting a TIA over conventional assemblies is that TIA do not require any adhesives. This allows easier disassembly of the structure once it reaches the end of its life cycle. Further, as the blocks in the assembly are only kinematically constrained, the propagation of cracks is more limited \cite{khor2004mechanisms}. This also allows easier repairs of the assemblies, particularly as some unused blocks can be stored on site, due to all of them possibly having the same geometry. Another advantage is the accessibility of the construction site. As the blocks are significantly smaller than the final structure, only smaller access to the building site is required.

Moreover, using interlocking blocks can minimize material usage, as demonstrated by the concrete blocks constructed in references \cite{gomes2025topological} and \cite{neef2024materialsparende}.

\subsection{Literature review}

The concept of topological interlocking assemblies has a long history. Abeille and Truchet first introduced interlocking assemblies \cite{gallon1735machines}, which were later generalized by Frézier \cite{frezier1738theorie}. The modern study of TIAs was initiated by Dyskin et al. \cite{dyskin2001toughening}, who established it as a novel material design concept. Their work demonstrated, for example, that all Platonic solids can be arranged as TIAs \cite{dyskin2003topological}. Further, the authors of \cite{dyskin2001new} have shown quite favourable responses of TIAs to local impacts.

% Figure 2: Blocks (a,c,e) and assemblies (b,d,f), the frame colored red.

More recently, a more general method for constructing planar TIAs based on wallpaper symmetries has been introduced in \cite{goertzen2024constructing,goertzen2022topological}. This approach was used to design e.g., the so-called Versatile Block, which admits three different symmetric arrangements.

The creation of large numbers of TIAs is facilitated by the new approach. As a relatively recent development in the field, only a limited number of research results are currently available. For instance, it has been shown that different arrangements of the same block can have significantly varying mechanical performances \cite{goertzen2025influence}. Thus, a crucial next step is to analyze the mechanical performance of the geometry of the blocks created using the new method, which will be presented here. Combined, the results of \cite{goertzen2025influence} with the present work reveal a large design space and demonstrate that both, arrangement and geometry variation, have a significant influence on the mechanical performance of a TIA. Moreover, they underscore that there is a high potential for optimization and custom tailoring of TIAs.

\section{Construction of the Blocks}

This section describes the construction of the starfish block, which is based on the method presented in \cite{goertzen2024constructing,goertzen2022topological}. The authors of \cite{goertzen2024constructing,goertzen2022topological} exploit a given wallpaper group to construct blocks admitting TIAs. Such a group yields a tiling of the plane consisting of copies of the same tile, called a \emph{fundamental domain}. The method begins by deforming a fundamental domain such that the result is another fundamental domain of the same wallpaper group. The two resulting fundamental domains are then placed on parallel planes and interpolating between them results in a block, respecting the chosen wallpaper group.

\begin{figure}[H]
    \centering
    \includegraphics[width=0.6\linewidth]{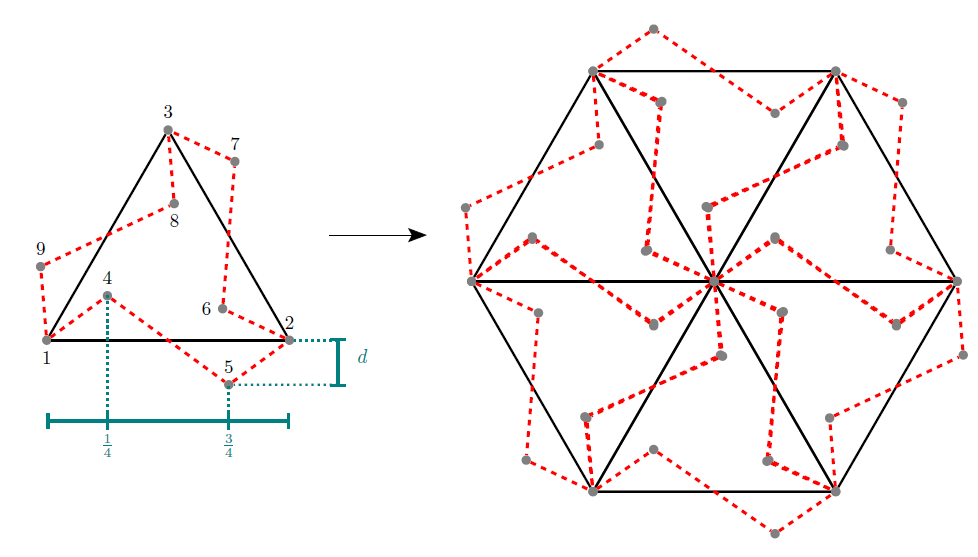}
    \caption{Deformation of the fundamental domain and a part of an assembly}
    \label{fig:fund-dom}
\end{figure}

Here we consider the wallpaper group $p6$, which is generated by a $60^\circ$ rotation around the origin together with two linearly independent translations. This group admits a hexagonal tiling of the plane, hence an equilateral triangle can form a fundamental domain (see black outline Fig. \ref{fig:fund-dom}).

\begin{figure}[H]
    \centering
    \includegraphics[width=0.6\linewidth]{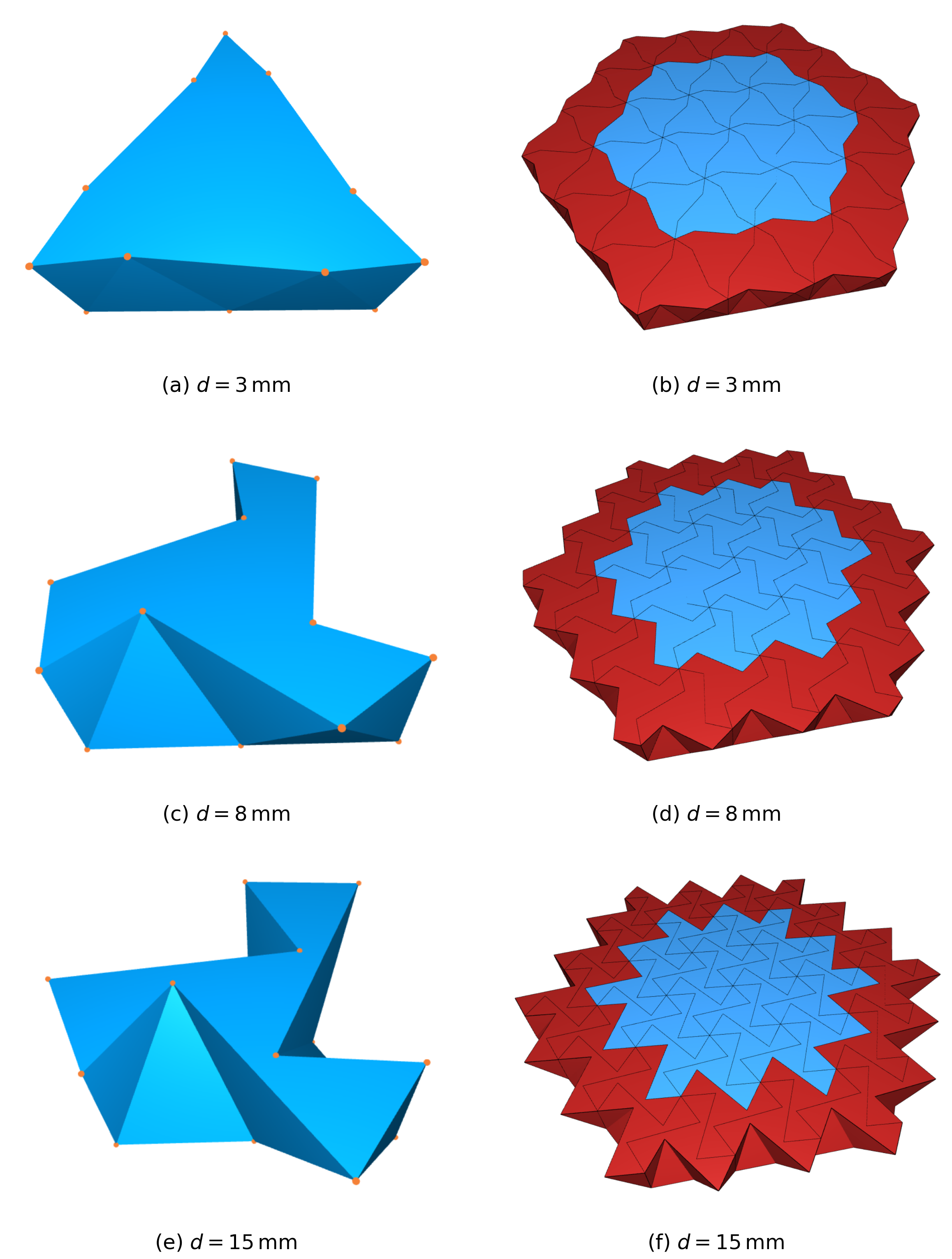}
    \caption{Blocks (a,c,e) and assemblies (b,d,f), the frame colored red.}
    \label{fig:blocks-assemblies}
\end{figure}

All edges of this triangle are then deformed by translating the points at $\frac{1}{4}$ and $\frac{3}{4}$ of the considered edge into the normal direction of the edge (see red outline Fig. \ref{fig:fund-dom}). The magnitude of this translation forms our proposed design parameter (denoted by $d$), which varies throughout our analysis. This deformation is depicted in Fig. \ref{fig:fund-dom} and results in another fundamental domain of the wallpaper group $p6$. As described, these two fundamental domains are placed on parallel planes which are $20 \mathrm{mm}$ apart and then interpolated between them, resulting in the so-called \emph{starfish block}. For a selection of parameter $d$, the blocks and their assemblies are shown in \ref{fig:blocks-assemblies}.

To achieve comparability, the triangle was chosen such that it has an area of $800 \mathrm{mm}^2$ resulting in an edge length of $a \approx 42.9829 \mathrm{mm}$. Thereby the starfish block considered here has the same height and volume as the Versatile Block in \cite{goertzen2025influence}.

\section{Simulation setup}

We consider blocks with parameter  $d$ ranging from $1 \mathrm{mm}$ to $11 \mathrm{mm}$ in $1 \mathrm{mm}$ increments as well as $15 \mathrm{mm}$, and $18.5 \mathrm{mm}$. As the largest possible value for $d$ without self-intersections is $x \coloneqq \frac{\sqrt{3}a}{4} \approx 18.6 \mathrm{mm}$, blocks with $d \geq x$ are not considered.

\subsection{Assembly generation}

The assemblies are generated using the CAD software Fusion 360. To take the design parameter $d$ into account, a user parameter is introduced. Then an initial starfish block is constructed and arranged into a planar assembly in accordance to the symmetry group $p6$. This assembly contains $24$ inner blocks and a frame consisting of the $30$ blocks on the perimeter. Next, the blocks are moved, so that there is a gap of $ 1 \times 10^{-5} \mathrm{mm}$ between all blocks which are in contact. The user parameter $d$ was varied to the chosen values, with the entire assembly exported as a .step file.

\subsection{Simulation setup}

The mechanical performance of the assembled blocks is evaluated using the commercial finite element software Abaqus/Explicit 2024 in a quasi-static simulation. The chosen material is steel with a Young's modulus of $210 \mathrm{GPa}$, a Poisson's ratio of $0.3$ and a density of $7.85 \frac{\mathrm{g}}{\mathrm{cm}^3}$. The contact formulation is surface-to-surface contact with a low friction of $0.01$ and an exponential pressure-overclosure relationship starting at $1 \times 10^{-4} \mathrm{mm}$ resulting in slight contact between the neighboring blocks due to the chosen gap.

The blocks in the frame are kinematically constrained and a uniform pressure load of $0.15 \mathrm{MPa}$ is applied to the side of the assembly consisting of the deformed fundamental domains. The length of the simulation is chosen as $2 \mathrm{s}$ to achieve kinematic equilibrium as the load is applied with smooth steps until $1 \mathrm{s}$.

After the simulations are finished, the maximum Mises stress, contact pressure as well as minimum deformation in direction of the applied load are extracted over the entire assembly.

\section{Results and discussion}
\subsection{Results}

\begin{figure}[H]
    \centering
    \includegraphics[width=0.5\linewidth]{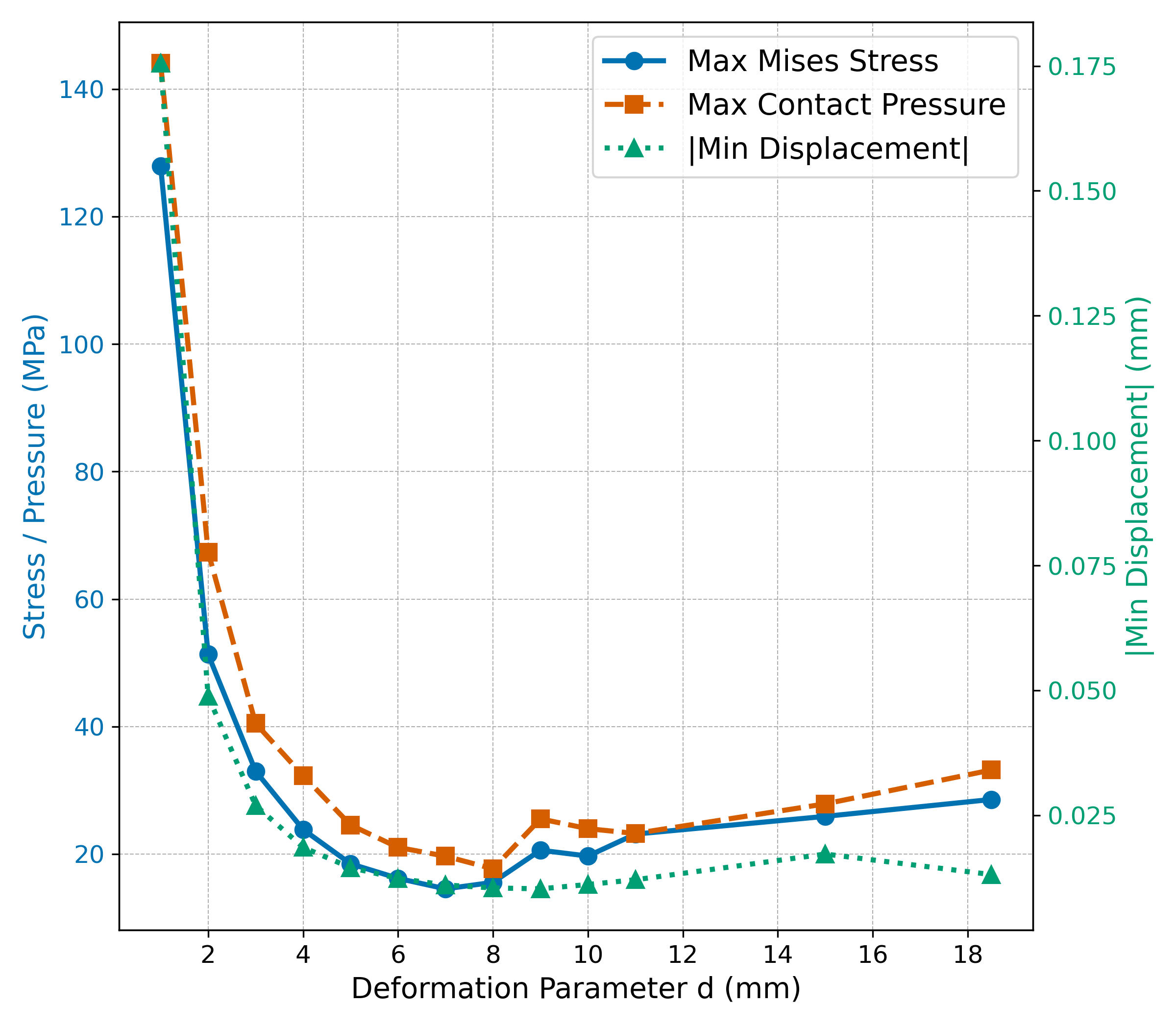}
    \caption{Parameter $d$ w.r.t. to mechanical performance indicators of the TIA.}
    \label{fig:curves}
\end{figure}

Three key mechanical performance indicators are chosen: the maximum Mises stress, the maximum contact pressure and the minimum deformation, in the direction in which the load is applied, all of them over the entire assembly. The relationship between the considered parameter $d$ and the performance indicators is summarized in Fig. \ref{fig:curves}.

All three indicators show that the mechanical performance improves significantly until around $d=10 \mathrm{mm}$. This suggests that increasing the parameter $d$ initially improves the mechanical performance, as the parts of the blocks that are kinematically constraining each increase in size, which leads to a better load distribution and thus reduces stress concentrations. This observation is also in good agreement with the fact that for $d=0 \mathrm{mm}$ the assembly does not admit a topological interlocking.

However, further increasing the parameter $d$, seems to reverse this trend. From around $d=10 \mathrm{mm}$ onwards the mechanical performance in all three indicators reverses, although it decreases not as much as the performance indicators of $d=1 \mathrm{mm}$, see Fig. \ref{fig:curves}.

\begin{figure}[H]
    \centering
    \includegraphics[width=0.6\linewidth]{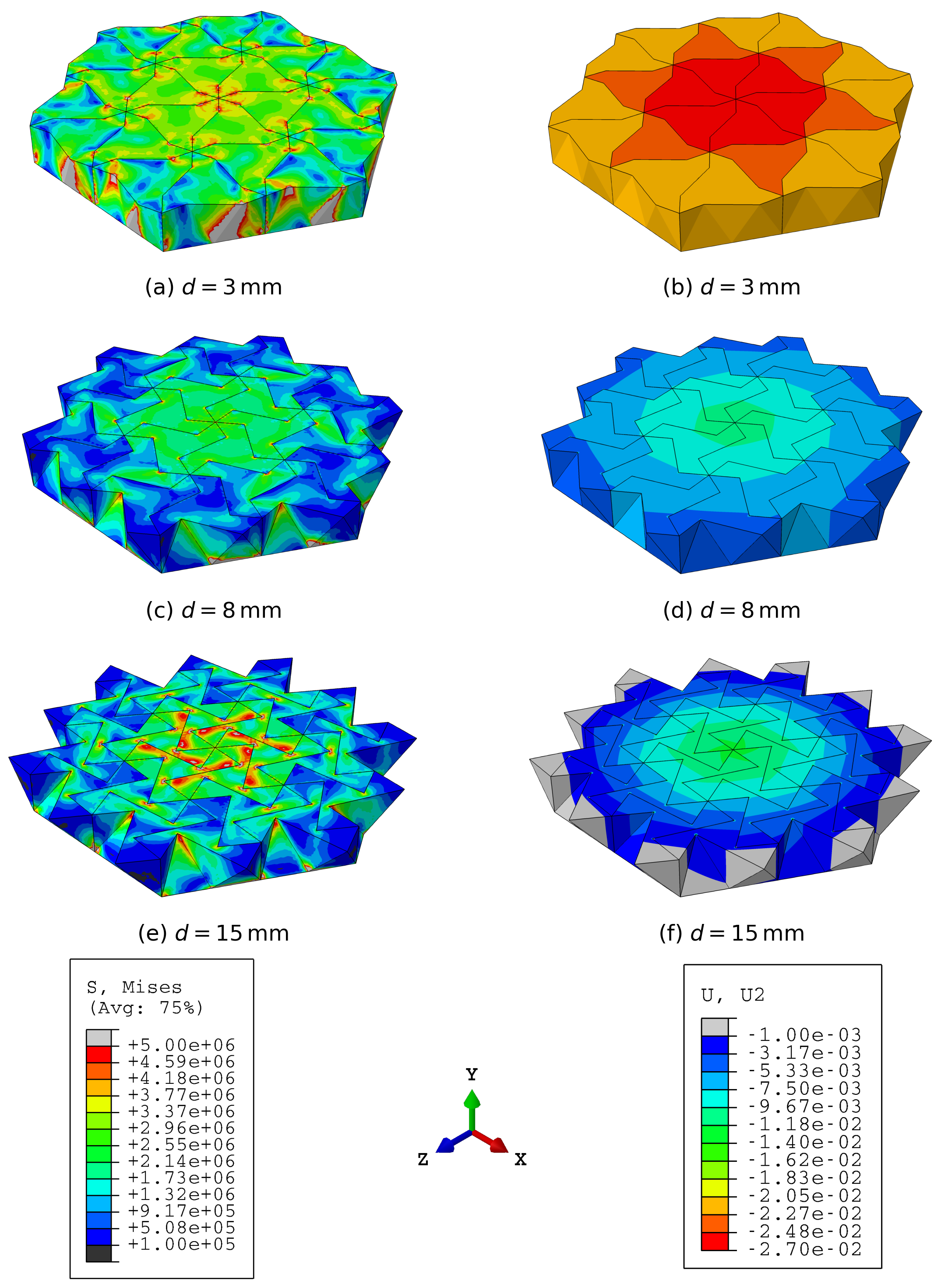}
    \caption{Stresses ($\mathrm{Pa}$) and deformations ($\mathrm{mm}$) on selected assemblies}
    \label{fig:fem-results}
\end{figure}

Fig. \ref{fig:fem-results} shows the FEM results for the Mises stress (left column) and the deformation (right column) of the entire assembly for $d = 3, 8,$ and $15 \mathrm{mm}$. Note that the frame is not shown in order to better highlight the internal results and the $U2$ direction is normal to the assembly plane. The maximal Mises stress and the maximal absolute deformation vary with the choice of parameter $d$, and the corresponding values in combination with the contact pressure are listed in Table \ref{tab:mech-results}. However, the distribution of the two mechanical performance indicators within the assemblies is similar.

\begin{table}[H]
    \centering
    \begin{tabular}{ | c | m{7em} | m{7em} | m{11em} | }
    \hline
    Parameter $d$ & Max.~Mises Stress [MPa] & Max.~contact pressure [MPa] & Max.~abs.~deformation normal to pressure [$\mu \mathrm{m}$] \\ 
    \hline
    $3 \mathrm{mm}$ & $33.00$ & $40.49$ & $26.95$ \\
    $8 \mathrm{mm}$ & $15.56$ & $17.69$ & $10.44$ \\
    $15 \mathrm{mm}$ & $25.91$ & $27.87$ & $17.23$\\
    \hline     
    \end{tabular}
    \caption{Some of the mechanical performance results.}
    \label{tab:mech-results}
\end{table}

\subsection{Discussion}

As noted above, mechanical performance increases up to $10 \mathrm{mm}$, after which the trend reverses. We hypothesize that this reversal is caused by thin regions and acute angles in the block resulting from the large deformation parameter moving the effected vertex close to another edge. These thin regions result in stress concentrations as the cone in which the stress can be transferred from one block onto a neighboring block cannot distribute the load over much material. An example of this stress concentration behavior for $d=15 \mathrm{mm}$ can be observed in Fig. \ref{fig:fem-results}e, where the points moved close to another edge show stress spikes, which reduce performance.

Our findings demonstrate the high level of influence the parameter $d$ has on the mechanical response of the assembly. Further, it seems that there is an optimum performance w.r.t. this design parameter, which is not at the boundaries of the possible values for $d$. Thus, an optimization of the parameter $d$ is necessary for maximizing the performance of the assembly. This optimization needs to balance the advantages of larger deformations, which distribute the load better, with the disadvantages of concentrated stresses from localized weaknesses.

\section{Conclusion and outlook}

The work presented in this paper shows that a design parameter for a block geometry can have a significant influence on the mechanical performance of a resulting TIA and particularly highlights that the optimum of the design parameter is not at the boundaries of its possible values. Combined with the findings of \cite{goertzen2025influence} this shows that the mechanical performance of TIAs can be tailored through different design parameters, both on the level of arrangements as well as on the level of the block geometry. These results provide a clearer understanding of how geometry affects mechanical performance of TIAs. Nevertheless, further investigations of the influence of block geometry are required, with emphasis on how specific geometric parameters influence load transfer, contact pressure, and stability, in order to achieve a comprehensive understanding of these relationships.

One significant limitation of our study is the limited focus on a single parameter used in the generation of the block. Future research should explore the influence of other parameters, such as the height or smoothing of the block, to understand their individual influence on the mechanical performance better. Additionally, the interplay between multiple parameters e.g. the design parameter $d$ and the height of the blocks, should be investigated, ideally revealing some scale-free parameters.

Furthermore, the current analysis was conducted on a relatively small assembly. Expanding the study to a larger-scale assembly could provide valuable insights into its behavior in a more general setting, and would also allow an accurate comparison with the results presented in \cite{goertzen2025influence}.

A similar parameter study could also be performed on other blocks derived from wallpaper groups, such as the Versatile Block or blocks with a smooth deformation. This could lead to a more holistic understanding of the influence in mechanical response from the underlying symmetry.

Lastly, the FEM computations took quite a long time resulting in considering only the presented load case even though TIAs are rigid assemblies for every load case. Thus, it could be advantageous, to develop a tailored tool for assemblies like the ones presented here. Such a tool would also aid the proposed optimization of the blocks.

\subsection*{Acknowledgements}

The authors gratefully acknowledge the funding by the Deutsche Forschungsgemeinschaft (DFG, German Research Foundation) in the framework of the Collaborative Research Centre CRC/TRR 280 ``Design Strategies for Material-Minimized Carbon Reinforced Concrete Structures--Principles of a New Approach to Construction'' (project ID 417002380).\\
We thank the referees for their helpful suggestions.

\subsection*{Data availability}

All relevant files can be found here:
\href{https://doi.org/10.5281/zenodo.17045363}{doi.org/10.5281/zenodo.17045363}

\nocite{*}
\printbibliography[heading=bibintoc,title={References}]

@inproceedings{khor2004mechanisms,
  author    = {Khor, H. C. and Dyskin, A. V. and Estrin, Y. and Pasternak, E.},
  title     = {Mechanisms of fracturing in structures built from topologically interlocked blocks},
  booktitle = {Structural Integrity and Fracture International Conference (SIF'04)},
  year      = {2004},
  venue     = {Brisbane, Australia},
  eventdate = {2004-09-26/2004-09-29}
}

@inproceedings{gomes2025topological,
  author    = {Gomes, C. G. and St\"{u}ttgen, S. and Wei\ss, M. and Akpanya, R. and Niemeyer, A. C. and Robertz, D. and Chudoba, R.},
  title     = {Topological interlocking assemblies based on origami inspired carbon-reinforced concrete waterbomb modules},
  booktitle = {fib Symposium 2025},
  year      = {2025}
}

@article{neef2024materialsparende,
  author    = {Neef, T. and Goertzen, T. and Niemeyer, A. C. and Mechtcherine, V.},
  title     = {Materialsparende Betondecke aus 3D-gedruckten Verriegelungsbl\"{o}cken},
  journal   = {Beton- und Stahlbetonbau},
  year      = {2024},
  volume    = {119},
  pages     = {882--893}
}

@book{gallon1735machines,
  author    = {Gallon, J.-G.},
  title     = {Machines et inventions approuv\'{e}es par l'Acad\'{e}mie Royale des Sciences depuis son \'{e}tablissement jusqu'\`{a} present; avec leur Description},
  publisher = {Acad\'{e}mie royale des Sciences},
  address   = {Paris},
  year      = {1735}
}

@book{frezier1738theorie,
  author    = {Fr\'{e}zier, A. F.},
  title     = {La theorie et la pratique de la coupe des pierres et des bois, pour la construction des voutes et autres parties des b\^{a}timens civils \& militaires, ou Trait\'{e} de stereotomie a l'usage de l'architecture},
  volume    = {2},
  publisher = {Doulsseker},
  address   = {Paris},
  year      = {1738}
}

@article{dyskin2001toughening,
  author    = {Dyskin, A. V. and Estrin, Y. and Kanel-Belov, A. J. and Pasternak, E.},
  title     = {Toughening by Fragmentation---How Topology Helps},
  journal   = {Advanced Engineering Materials},
  year      = {2001},
  volume    = {3},
  number    = {11},
  pages     = {885--888}
}

@article{dyskin2003topological,
  author    = {Dyskin, A. V. and Estrin, Y. and Kanel-Belov, A. J. and Pasternak, E.},
  title     = {Topological interlocking of platonic solids: A way to new materials and structures},
  journal   = {Philosophical Magazine Letters},
  year      = {2003},
  month     = jan,
  volume    = {83},
  number    = {3},
  pages     = {197--203}
}

@article{dyskin2001new,
  author    = {Dyskin, A. V. and Estrin, Y. and Kanel-Belov, A. J. and Pasternak, E.},
  title     = {A new concept in design of materials and structures: assemblies of interlocked tetrahedron-shaped elements},
  journal   = {Scripta Materialia},
  year      = {2001},
  volume    = {44},
  number    = {12},
  pages     = {2689--2694}
}

@phdthesis{goertzen2024constructing,
  author    = {Goertzen, T.},
  title     = {Constructing simplicial surfaces with given geometric constraints},
  school    = {RWTH Aachen University},
  address   = {Aachen, Germany},
  year      = {2024},
  type      = {Dissertation},
  note      = {Report No.: RWTH-2024-08038},
  doi       = {10.18154/RWTH-2024-08038}
}

@inproceedings{goertzen2022topological,
  author    = {Goertzen, T. and Niemeyer, A. C. and Plesken, W.},
  title     = {Topological Interlocking via Symmetry},
  booktitle = {6th fib International Congress on Concrete Innovation for Sustainability},
  editor    = {Stokkeland, S. and Braarud, H. C.},
  year      = {2022},
  pages     = {1}
}

@article{goertzen2025influence,
  author    = {Goertzen, T. and Macek, D. and Schnelle, L. and Wei\ss, M. and Reese, S. and Holthusen, H. and Niemeyer, A. C.},
  title     = {Influence of block arrangement on mechanical performance in topological interlocking assemblies: A study of the versatile block},
  journal   = {International Journal of Solids and Structures},
  year      = {2025},
  volume    = {306},
  pages     = {113102}
}

\end{document}